# A tale of two databases: The use of Web of Science and Scopus in academic papers

Junwen Zhu
jwzhu@ed.ecnu.edu.cn
Faculty of Education, East China Normal University, Shanghai 200062, China

Weishu Liu Corresponding author
wsliu08@163.com
School of Information Management and Engineering, Zhejiang University of Finance and Economics, Hangzhou 310018, Zhejiang, China

**Abstract** Web of Science and Scopus are two world-leading and competing citation databases. By using the Science Citation Index Expanded and Social Sciences Citation Index, this paper conducts a comparative, dynamic, and empirical study focusing on the use of Web of Science (WoS) and Scopus in academic papers published during 2004 and 2018. This brief communication reveals that although both Web of Science and Scopus are increasingly used in academic papers, Scopus as a new-comer is really challenging the dominating role of WoS. Researchers from more and more countries/regions and knowledge domains are involved in the use of these two databases. Even though the main producers of related papers are developed economies, some developing economies such as China, Brazil and Iran also act important roles but with different patterns in the use of these two databases. Both two databases are widely used in meta-analysis related studies especially for researchers in China. Health/medical science related domains and the traditional Information Science & Library Science field stand out in the use of citation databases.

## Introduction

As the most important legacy of Eugene Garfield (Li et al. 2018; Jacso 2018), Web of Science (WoS) Core Collection especially its three classical journal citation indexes, i.e. Science Citation Index Expanded (SCIE), Social Sciences Citation Index (SSCI), and Arts & Humanities Citation Index (A&HCI), are well-known and widely used in academia (Hu et al. 2018; Liu et al. 2020; Tang & Shapira 2011). By focusing on the database itself, Li et al. (2018) conduct a pioneer empirical analysis on the use of Web of Science during 1997 and 2017 and uncover the characteristics of the academic use of WoS across countries/regions, institutions, and knowledge domains. Moreover, in order to depict the non-transparent use of WoS, Liu (2019) also finds that an increasing number of papers have mentioned WoS in their topic field.

Although the new-comer Scopus was launched in 2004, it is a powerful competitor of WoS and is attempting to challenge the dominating role of WoS. Various studies have compared these two databases from different perspectives (Abdulhayoglu & Thijs 2018; Adriaanse & Rensleigh 2013; Harzing & Alakangas 2016; Martín-Martín et al. 2018; Meho & Sugimoto 2009; Moed et al. 2018; Mongeon & Paul-Hus 2016; Wang & Waltman 2016; Zhu et al. 2019a, 2019b). However, to the best of our knowledge, no empirical study has been conducted focusing on the use of Scopus in academic papers let alone a comparative study about both of them. Some questions are interesting for further investigation: 1) Is Scopus really threatening the dominating role of WoS? 2) Do the researchers from different countries/regions and research fields have any preference in choosing these two databases?

This study tries to answer these questions by conducting a comparative, dynamic, and empirical analysis focusing on the use of WoS and Scopus in academic papers. The remaining part of this paper is organized as follows. This study first describes the data and methods used in this research and then presents the dynamics, main contributors and knowledge domains of the use of WoS and Scopus in academic research respectively. Lastly, this study ends with the conclusion and discussion.

## Data and methods

The web-based Web of Science was launched in 1997 and renamed Web of Science Core Collection around 2014[1]. The WoS integrated SCIE, SSCI and A&HCI indexes initially in 1997[2] and expanded its coverage gradually (Liu 2019; Rousseau et al. 2018). To keep consistent, this study uses "Web of Science" and the abovementioned three index names as the keywords to retrieve WoS-related records[3]. The Scopus database was launched in 2004, therefore this study sets the time span between 2004 and 2018 for analysis. The WoS's topic field is used to search via the advanced search platform[4]. The following two queries are used to search WoS- and Scopus-related records[5]. The data source is limited to SCIE and SSCI. The search was conducted on 9th August via the library of

[1] http://wokinfo.com/nextgenwebofscience?elq=4e2a3b0638fb400cae0565fc0e03a24e&elqCampaignId=8201

[2] https://www.thomsonreuters.cn/zh/about-us/company-history.html

[3] The search strategy used in this study is a bit different to that used by Li et al. (2018), both these two search strategies may introduce a very small percentage of records which have only mentioned some regional citation indexes such as Chinese Social Sciences Citation Index.

[4] Although Web of Science's topic search (search in title, abstract, author keywords and keywords plus fields) is widely used in practice, the search in the keywords plus field may introduce some noise. Besides, records which only mention the data sources in the data and methods section will also be omitted in this study.

[5] According to Wikipedia, Scopus also has some other meanings. This study excluded these ambiguous records manually.

Shanghai Jiao Tong University.

#1 TS=("Web of Science" OR "Science Citation Index" OR "Social Science* Citation Index" OR "Art* and Humanit* Citation Index" OR "Art* & Humanit* Citation Index")

#2 TS="Scopus"

Indexes=SCIE, SSCI; Timespan=2004-2018

## Analyses

### Dynamics of the use of WoS and Scopus in academic papers

The search query #1 retrieves 22890 hits of WoS-related records. This study keeps 22648 articles and reviews for further analysis. Figure 1 depicts the dynamics of the annual production of WoS-related papers. The number of papers mentioning "WoS" rose rapidly from 102 in 2004 to 4932 in 2018, especially after 2011, which demonstrates the increasing use of WoS in scientific papers.

After manually excluding 18 ambiguous records, the study identifies 12953 Scopus-related records published during 2004 and 2018 in SCIE and SSCI databases. 12861 articles and reviews are selected for further analysis. Similar to the WoS-related records, the annual production of Scopus-related records also grows rapidly. Only two Scopus-related papers were published in 2005[6], however, the number of Scopus-related papers went up to 3252 in 2018. As a new database, Scopus is increasingly used (at least mentioned) in academic papers (only a bit less than the competitor WoS) and is challenging the dominating role of WoS, which is obviously revealed by Figure 1.

[6]Two records published in 2004 are related to Scopus, however, one of them is news item and another one is editorial material which are excluded from this study.

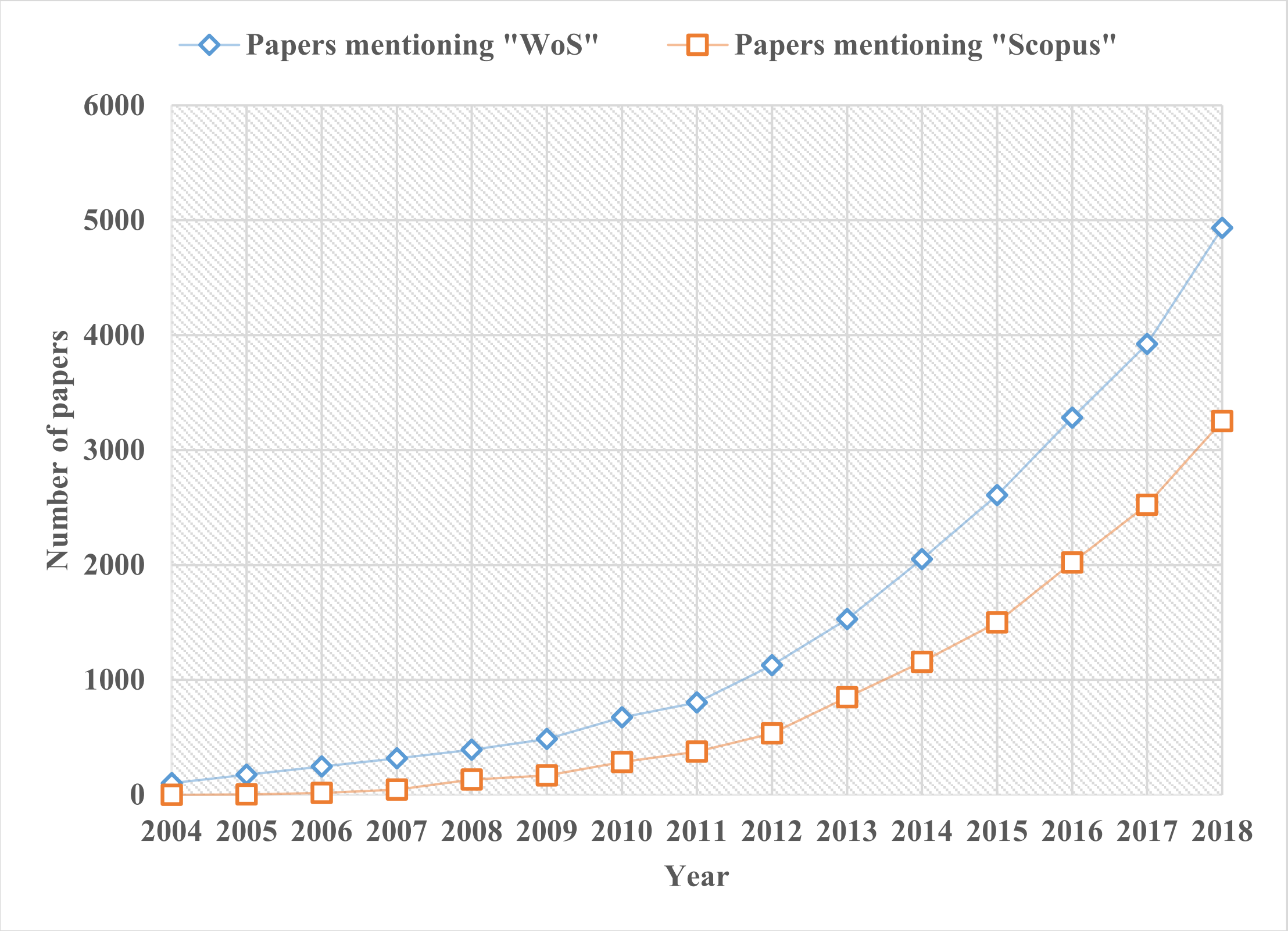

Figure 1 Dynamics of WoS- and Scopus-related papers

Note: Only articles and reviews are considered.

**Main contributors of WoS- and Scopus-related papers**

During the past 15 years, researchers from over 140 countries/regions have contributed to WoS-related papers. China leads with 6938 (30.6%) papers, followed by the USA (4261, 18.8%) and the UK (3372, 14.9%)[7]. Similarly, about 140 countries/regions have contributed to Scopus-related papers. Comparatively, the USA leads with 3093 (24.0%) Scopus-related papers followed by the UK (1590, 12.4%) and Australia (1438, 11.2%). However, China, although the largest contributor of WoS-related papers, only ranks as the 9th contributor of Scopus-related papers (737, 5.7%).

In order to depict the dynamics of the main contributors of WoS- and Scopus-related papers, this study splits the 15-year period into three successive 5-year phases: 2004-2008, 2009-2013, and 2014-2018. Table 1 lists the top 10 countries/regions which have contributed most to WoS- and Scopus-related research in each phase.

The USA, as the dominating research power, takes the lead in the number of Scopus-related papers in all the three phases. Researchers from the USA also produced the largest number of WoS-related papers during the first two phases but were replaced by China in the third phase. Although the number of related papers produced by the USA is increasing during all the three phases, its relative share is decreasing for both WoS- and Scopus-related papers.

[7]Echoing the finding of Liu et al. (2017) and Liu et al. (2018), a small percentage of country/region information omission is also identified. This study merges England, Scotland, Wales and North Ireland into the UK.

According to Table 1, most of the main contributors of these two databases related papers are also developed economies. One possible explanation is that these developed economies have enough budget to subscribe to these two expensive databases while the budget of many developing economies may be limited. However, Brazil, China and Iran, the three developing economies are also the main contributors of related research. What's more, they demonstrate different patterns regarding the publishing of WoS- and Scopus-related papers. Researchers from Brazil contribute 1202 (5.3%, 7th) WoS-related papers and 1130 (8.8%, 5th) Scopus-related papers during the past 15 years. However, as mentioned before, researchers from China produced much more WoS-related papers than Scopus-related papers among all the three phases (from both absolute and relative perspectives). Contrarily, Iran contributed 670 (3.0%, 11th) WoS-related papers but 1105 (8.6%, 6th) Scopus-related papers.

Table 1 Main contributors of WoS-and Scopus-related papers

| Phase | | Papers mentioning Web of Science | | | Papers mentioning Scopus | | |
|---|---|---|---|---|---|---|---|
| | Rank | Countries/regions | # | % | Countries/regions | # | % |
| 2004-2008 | 1 | USA | 339 | 27.6 | USA | 71 | 36.6 |
| | 2 | UK | 231 | 18.8 | UK | 28 | 14.4 |
| | 3 | Canada | 112 | 9.1 | Canada | 24 | 12.4 |
| | 4 | Netherlands | 92 | 7.5 | Greece | 24 | 12.4 |
| | 5 | Spain | 82 | 6.7 | Germany | 16 | 8.2 |
| | 6 | Australia | 60 | 4.9 | Iran | 12 | 6.2 |
| | 7 | Peoples R China | 51 | 4.1 | Italy | 12 | 6.2 |
| | 8 | Germany | 49 | 4.0 | Switzerland | 7 | 3.6 |
| | 9 | Brazil | 43 | 3.5 | Israel | 5 | 2.6 |
| | 10 | Denmark | 42 | 3.4 | Netherlands | 5 | 2.6 |
| 2009-2013 | 1 | USA | 1012 | 21.9 | USA | 656 | 29.7 |
| | 2 | Peoples R China | 925 | 20.0 | UK | 270 | 12.2 |
| | 3 | UK | 895 | 19.4 | Australia | 208 | 9.4 |
| | 4 | Canada | 397 | 8.6 | Canada | 199 | 9.0 |
| | 5 | Netherlands | 338 | 7.3 | Brazil | 155 | 7.0 |
| | 6 | Australia | 317 | 6.9 | Spain | 153 | 6.9 |
| | 7 | Spain | 242 | 5.2 | Italy | 145 | 6.6 |
| | 8 | Germany | 205 | 4.4 | Iran | 133 | 6.0 |
| | 9 | Brazil | 202 | 4.4 | Greece | 118 | 5.3 |
| | 10 | Italy | 191 | 4.1 | Netherlands | 118 | 5.3 |
| 2014- | 1 | Peoples R China | 5962 | 35.5 | USA | 2366 | 22.6 |
| | 2 | USA | 2910 | 17.3 | UK | 1292 | 12.4 |

| | | | | | | |
|---|---|---|---|---|---|---|
| 3 | UK | 2246 | 13.4 | Australia | 1228 | 11.7 |
| 4 | Australia | 1149 | 6.8 | Italy | 1084 | 10.4 |
| 5 | Canada | 1028 | 6.1 | Brazil | 973 | 9.3 |
| 6 | Brazil | 957 | 5.7 | Iran | 960 | 9.2 |
| 7 | Netherlands | 807 | 4.8 | Spain | 685 | 6.6 |
| 8 | Italy | 787 | 4.7 | Canada | 684 | 6.5 |
| 9 | Germany | 701 | 4.2 | Peoples R China | 644 | 6.2 |
| 10 | Spain | 697 | 4.1 | Germany | 369 | 3.5 |

Note: #, number of papers; %, relative share. Only articles and reviews are considered.

**Knowledge domains of relevant papers**

***Role of meta-analysis***

It is well known by researchers in the field of library and information science that both the classical Web of Science and the rising star Scopus are widely used in bibliometric related studies (Ellegaard 2018; Lei & Liu 2019; Yu et al. 2018). By using the following search queries #3 and #4, this study also identifies both citation database and meta-analysis related records[8].

#3 TS=("Meta analy*" OR "Metaanaly*") and #1

#4 TS=("Meta analy*" OR "Metaanaly*") and #2

According to the data, 50.1% of the WoS-related records published during the past 15 years are also meta-analysis related. Comparatively, 39.1% of the Scopus-related records are meta-analysis related. Figure 2 demonstrates the increase of WoS+meta-analysis and Scopus+meta-analysis related records in SCIE and SSCI databases. That is to say, both WoS and Scopus are widely used in meta-analysis related studies. Another surprising finding is that 50.0% of all the WoS+meta-analysis related papers are contributed by China followed by the USA (15.8%). Comparatively, the USA still leads with 27.0% of Scopus+meta-analysis related papers and China ranks as the 5th contributor with the share of 10.3%.

[8] For information about meta-analysis, please refer to Gurevitch et al. (2018). A similar search method was also used by Guilera et al. (2013).

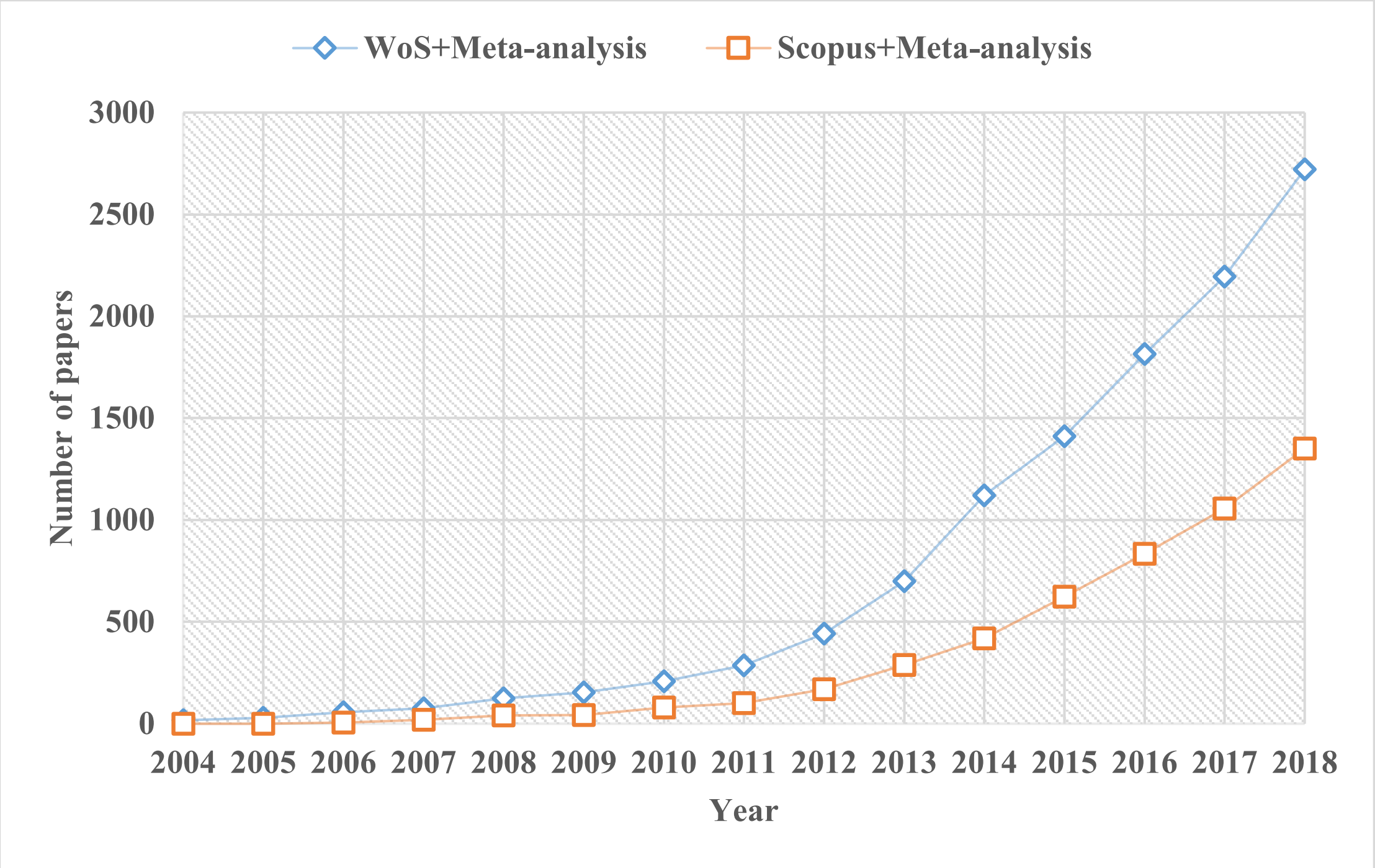

Figure 2 Dynamics of meta-analysis related papers

Note: Only articles and reviews are considered.

***Distribution of Web of Science categories***

The WoS-related papers published during the past 15 years cover over 200 Web of Science categories. The category Medicine, General & Internal leads with 3347 papers (14.8%), followed by Oncology (1692, 7.5%), Information Science & Library Science (1371, 6.1%), Public, Environmental & Occupational Health (1262, 5.6%) and Surgery (1236, 5.5%).

Similarly, the Scopus is also widely used (at least mentioned) in over 200 Web of Science categories. Medicine, General & Internal also leads with 1037 papers (8.1%), followed by Pharmacology & Pharmacy (914, 7.1%), Surgery (864, 6.7%), Public, Environmental & Occupational Health (856, 6.7%) and Information Science & Library Science (662, 5.1%).

The Science overlay maps of WoS- and Scopus-related papers during the whole 15-year study period are demonstrated in Figure 3 and Figure 4 (Leydesdorff et al. 2013; Zhang et al. 2016). The size of the node is positively associated with the number of papers in each category. The Science overlay map gives a full picture of WoS- and Scopus-related papers. As evidenced by the Figure 3 and Figure 4, both WoS and Scopus are widely used in various domains.

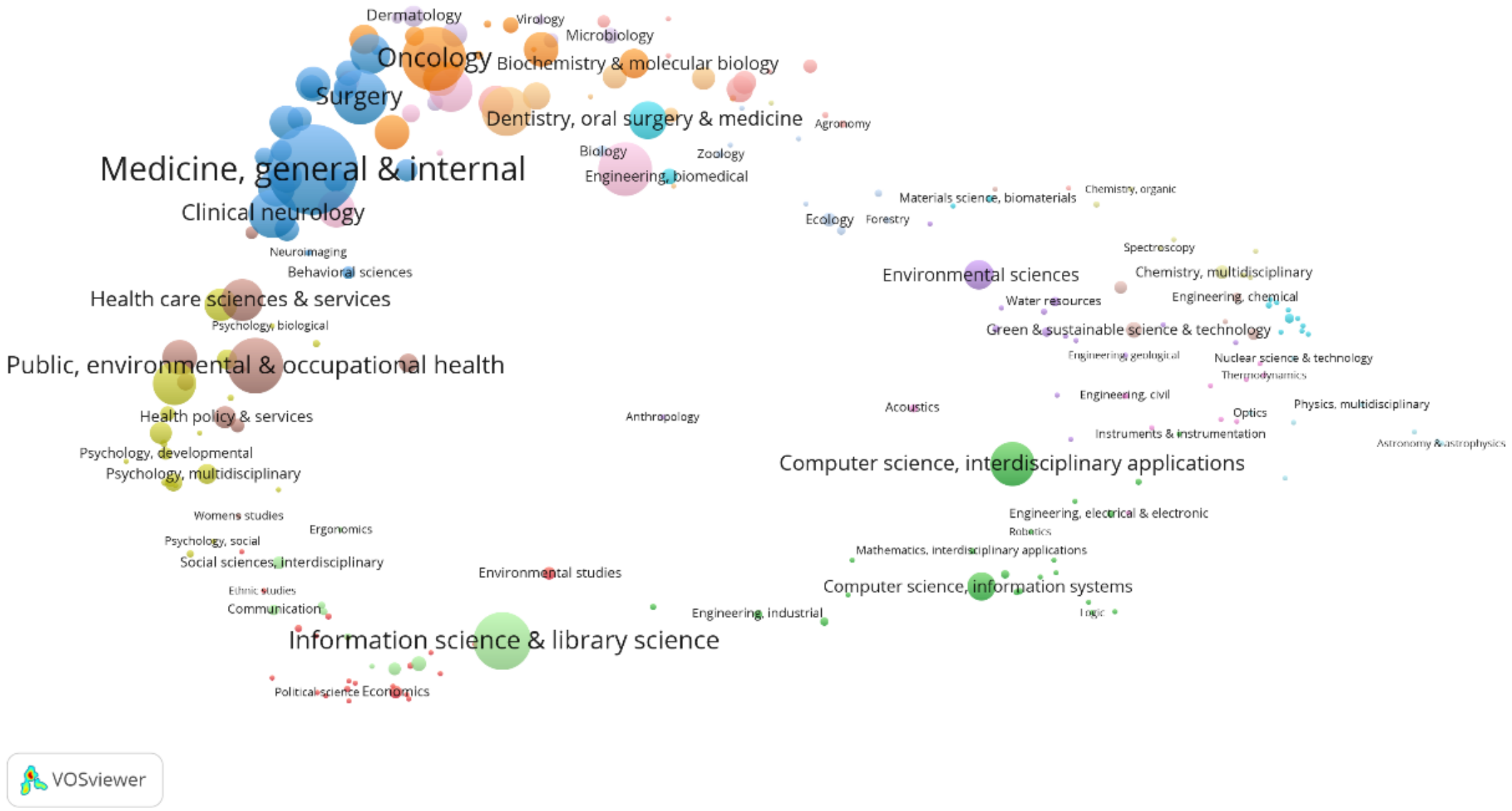

Figure 3 Science overlay map of WoS-related papers (2004-2018)

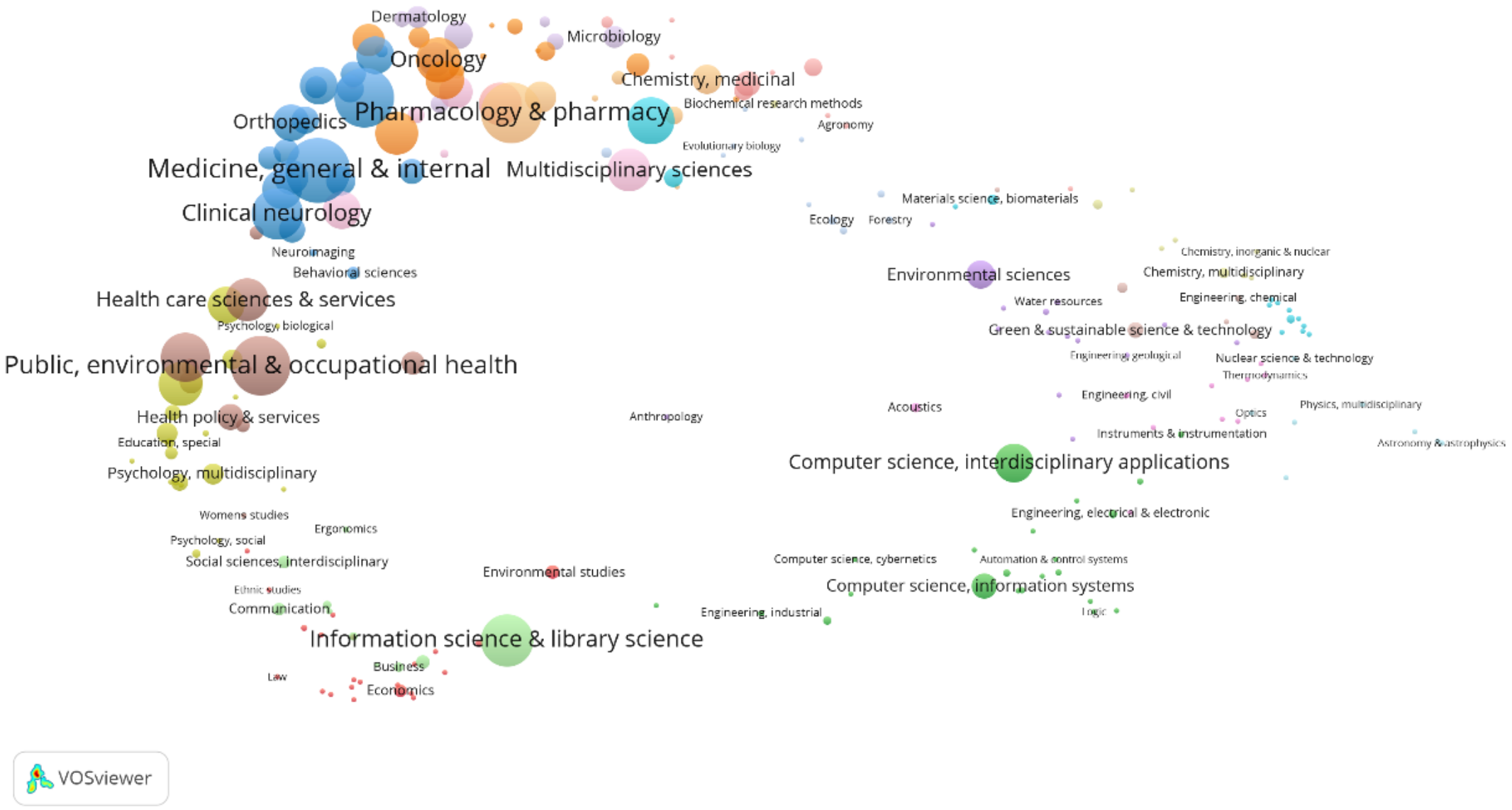

Figure 4 Science overlay map of Scopus-related papers (2004-2018)

In order to depict the dynamics of the distribution of Web of Science categories, the top 10 categories of each phase are listed in Table 2. We also calculate the relative share of each category among all the WoS/Scopus-related records and the ratio of studies that use WoS/Scopus in a specific category relative to the total number of studies in that category. Echoing the finding of Li et al. (2018) and Liu (2019) based on WoS-related papers, health/medical science related categories play important roles in taking WoS and Scopus as the data source for academic research. Information Science & Library Science also stands out. Although papers belong to this category also grow gradually, its rankings and relative shares (columns: % Within WoS/Scopus studies) decrease for both two groups[9]. However, compared to other top categories, the use of WoS/Scopus in this category are always more frequently for all the three phases (columns: % Within entire category). Besides, both WoS and Scopus are more and more frequently mentioned in Information Science & Library Science records evidenced by the rising shares provided by columns of % Within entire category in Table 2.

Medicine, General & Internal is the largest category for both groups. This category leads in all the three phases for both two groups, with the exception of the first phase in the Scopus group. However, along with the increasing number of categories involving in the use of WoS and Scopus for academic research, the relative share of papers in the category of Medicine, General & Internal is also decreasing for both two groups (columns: % Within WoS/Scopus studies). Besides, the relative shares of Medicine, General & Internal for the WoS group are much higher than the Scopus group for all the three phases (columns: % Within WoS/Scopus studies).

[9] The decease of relative shares (columns: % within WoS/Scopus studies) in Information Science & Library Science is due to faster growth rates in some other categories where literature mentions WoS/Scopus.

Table 2 Main categories of WoS- and Scopus-related papers

| Phase | Papers mentioning Web of Science | | | | | Papers mentioning Scopus | | | |
|---|---|---|---|---|---|---|---|---|---|
| | Rank | Web of Science Categories | # WoS studies | % Within WoS studies | % Within entire category | Web of Science Categories | # Scopus studies | % Within Scopus studies | % Within entire category |
| 2004-2008 | 1 | Medicine, General & Internal | 309 | 25.1 | 0.4 | Information Science & Library Science | 33 | 17.0 | 0.2 |
| | 2 | Information Science & Library Science | 193 | 15.7 | 1.4 | Medicine, General & Internal | 30 | 15.5 | 0.0 |
| | 3 | Computer Science, Interdisciplinary Applications | 90 | 7.3 | 0.2 | Pharmacology & Pharmacy | 21 | 10.8 | 0.0 |
| | 4 | Public, Environmental & Occupational Health | 74 | 6.0 | 0.1 | Clinical Neurology | 17 | 8.8 | 0.0 |
| | 5 | Computer Science, Information Systems | 67 | 5.4 | 0.1 | Anesthesiology | 16 | 8.2 | 0.1 |
| | 6 | Pharmacology & Pharmacy | 62 | 5.0 | 0.0 | Computer Science, Information Systems | 16 | 8.2 | 0.0 |
| | 7 | Psychiatry | 42 | 3.4 | 0.1 | Computer Science, Interdisciplinary Applications | 10 | 5.2 | 0.0 |
| | 8 | Surgery | 41 | 3.3 | 0.0 | Infectious Diseases | 9 | 4.6 | 0.0 |
| | 9 | Gastroenterology & Hepatology | 40 | 3.3 | 0.1 | Dentistry, Oral Surgery & Medicine | 8 | 4.1 | 0.0 |
| | 10 | Health Care Sciences & Services | 37 | 3.0 | 0.1 | Obstetrics & Gynecology | 8 | 4.1 | 0.0 |
| 2009-2013 | 1 | Medicine, General & Internal | 908 | 19.7 | 0.9 | Medicine, General & Internal | 204 | 9.2 | 0.2 |
| | 2 | Information Science & Library Science | 446 | 9.7 | 2.5 | Information Science & Library Science | 199 | 9.0 | 1.1 |
| | 3 | Oncology | 254 | 5.5 | 0.2 | Pharmacology & Pharmacy | 194 | 8.8 | 0.1 |
| | 4 | Computer Science, Interdisciplinary Applications | 245 | 5.3 | 0.4 | Public, Environmental & Occupational Health | 133 | 6.0 | 0.1 |
| | 5 | Public, Environmental & Occupational Health | 243 | 5.3 | 0.2 | Surgery | 112 | 5.1 | 0.1 |
| | 6 | Surgery | 211 | 4.6 | 0.1 | Computer Science, Interdisciplinary Applications | 96 | 4.3 | 0.2 |
| | 7 | Multidisciplinary Sciences | 208 | 4.5 | 0.1 | Nursing | 91 | 4.1 | 0.3 |
| | 8 | Pharmacology & Pharmacy | 196 | 4.2 | 0.1 | Clinical Neurology | 90 | 4.1 | 0.1 |
| | 9 | Health Care Sciences & Services | 183 | 4.0 | 0.5 | Oncology | 87 | 3.9 | 0.1 |
| | 10 | Psychiatry | 174 | 3.8 | 0.2 | Psychiatry | 84 | 3.8 | 0.1 |
| 2014-2018 | 1 | Medicine, General & Internal | 2130 | 12.7 | 1.7 | Medicine, General & Internal | 803 | 7.7 | 0.6 |
| | 2 | Oncology | 1410 | 8.4 | 0.6 | Surgery | 744 | 7.1 | 0.4 |
| | 3 | Surgery | 984 | 5.9 | 0.6 | Public, Environmental & Occupational Health | 716 | 6.8 | 0.5 |
| | 4 | Multidisciplinary Sciences | 969 | 5.8 | 0.3 | Pharmacology & Pharmacy | 699 | 6.7 | 0.3 |
| | 5 | Public, Environmental & Occupational Health | 945 | 5.6 | 0.6 | Clinical Neurology | 548 | 5.2 | 0.4 |
| | 6 | Clinical Neurology | 752 | 4.5 | 0.5 | Nursing | 485 | 4.6 | 1.2 |
| | 7 | Pharmacology & Pharmacy | 752 | 4.5 | 0.4 | Dentistry, Oral Surgery & Medicine | 471 | 4.5 | 1.0 |
| | 8 | Information Science & Library Science | 732 | 4.4 | 3.5 | Information Science & Library Science | 430 | 4.1 | 2.0 |
| | 9 | Medicine, Research & Experimental | 688 | 4.1 | 0.5 | Oncology | 418 | 4.0 | 0.2 |
| | 10 | Psychiatry | 560 | 3.3 | 0.6 | Nutrition & Dietetics | 402 | 3.8 | 0.6 |

Note: # WoS/Scopus studies, number of WoS/Scopus-related studies; % Within WoS/Scopus studies, relative share within all WoS/Scopus-related studies; % Within entire category, relative share within all SCI/SSCI records in this category. Only articles and reviews are considered.

### *Distribution of main publishing journals*

During the past 15 years, over 3000 journals have published WoS-related papers and over 2500 journals have published Scopus-related papers. Cochrane Database of Systematic Reviews leads with 1506 (6.7%) WoS-related papers, followed by PLoS One (904, 4.0%), Scientometrics (621, 2.7%), Medicine (498, 2.2%), and BMJ Open (354, 1.6%). Comparatively, PLoS One leads with 349 (2.7%) Scopus-related papers followed by Scientometrics (260, 2.0%), BMJ Open (189, 1.5%), Cochrane Database of Systematic Reviews (132, 1.0%), and Journal of Ethnopharmacology (101, 0.8%). These two groups share four common journals among each group's top 5 journals, although the rankings are a bit different. The relative shares of top journals for Scopus-related papers are much smaller than that for WoS-related papers, which indicates more even distribution of Scopus-related papers among publishing journals.

Table 3 lists the top 10 journals in each phase for two groups. Although more and more journals are involved in publishing WoS- or Scopus-related papers, most of the top journals are from the domain of library and information science or health/medical science. Although the main journals from the domain of health/medical science are a bit different for the two groups, they share some library and information science journals such as Scientometrics, Journal of the American Society for Information Science and Technology, and Journal of Informetrics.

Table 3 Main publishing journals of WoS- and Scopus-related papers

| Phase | Rank | Papers mentioning Web of Science Journals | # | % | Papers mentioning Scopus Journals | # | % |
|---|---|---|---|---|---|---|---|
| 2004-2008 | 1 | Cochrane Database of Systematic Reviews | 210 | 17.1 | Schmerz | 10 | 5.2 |
| | 2 | Scientometrics | 78 | 6.3 | Online Information Review | 7 | 3.6 |
| | 3 | Journal of the American Society for Information Science and Technology | 34 | 2.8 | Journal of the American Society for Information Science and Technology | 6 | 3.1 |
| | 4 | JAMA Journal of the American Medical Association | 17 | 1.4 | Clinical Therapeutics | 5 | 2.6 |
| | 5 | Annals of Pharmacotherapy | 15 | 1.2 | Journal of Informetrics | 5 | 2.6 |
| | 6 | Journal of Advanced Nursing | 11 | 0.9 | Scientometrics | 5 | 2.6 |
| | 7 | Annals of Internal Medicine | 10 | 0.8 | Cochrane Database of Systematic Reviews | 4 | 2.1 |
| | 8 | BMJ British Medical Journal | 9 | 0.7 | JAMA Journal of the American Medical Association | 4 | 2.1 |
| | 9 | Journal of Informetrics | 9 | 0.7 | Journal of Antimicrobial Chemotherapy | 4 | 2.1 |
| | 10 | Alimentary Pharmacology Therapeutics | 8 | 0.7 | Archives of Internal Medicine | 3 | 1.5 |
| 2009-2013 | 1 | Cochrane Database of Systematic Reviews | 612 | 13.2 | Scientometrics | 68 | 3.1 |
| | 2 | Scientometrics | 188 | 4.1 | PLoS One | 61 | 2.8 |
| | 3 | PLoS One | 182 | 3.9 | Cochrane Database of Systematic Reviews | 49 | 2.2 |
| | 4 | Journal of the American Society for Information Science and Technology | 70 | 1.5 | Journal of Informetrics | 24 | 1.1 |
| | 5 | Health Technology Assessment | 57 | 1.2 | Journal of the American Society for Information Science and Technology | 24 | 1.1 |
| | 6 | Journal of Informetrics | 46 | 1.0 | Journal of Advanced Nursing | 17 | 0.8 |
| | 7 | Asian Pacific Journal of Cancer Prevention | 32 | 0.7 | Annals of Internal Medicine | 15 | 0.7 |
| | 8 | BMJ British Medical Journal | 31 | 0.7 | European Urology | 15 | 0.7 |
| | 9 | Tumor Biology | 30 | 0.6 | Journal of Ethnopharmacology | 15 | 0.7 |
| | 10 | World Journal of Gastroenterology | 29 | 0.6 | Online Information Review | 15 | 0.7 |
| 2014-2018 | 1 | PLoS One | 718 | 4.3 | PLoS One | 288 | 2.8 |
| | 2 | Cochrane Database of Systematic Reviews | 684 | 4.1 | Scientometrics | 187 | 1.8 |
| | 3 | Medicine | 498 | 3.0 | BMJ Open | 180 | 1.7 |
| | 4 | Scientometrics | 355 | 2.1 | Journal of Ethnopharmacology | 85 | 0.8 |
| | 5 | BMJ Open | 337 | 2.0 | Cochrane Database of Systematic Reviews | 79 | 0.8 |
| | 6 | International Journal of Clinical and Experimental Medicine | 292 | 1.7 | Medicine | 67 | 0.6 |
| | 7 | Oncotarget | 282 | 1.7 | Sports Medicine | 59 | 0.6 |
| | 8 | Oncotargets and Therapy | 138 | 0.8 | International Journal of Nursing Studies | 56 | 0.5 |
| | 9 | Scientific Reports | 137 | 0.8 | Iranian Journal of Public Health | 54 | 0.5 |
| | 10 | Tumor Biology | 118 | 0.7 | Journal of Informetrics | 51 | 0.5 |

Note: #, number of papers; %, relative share. Only articles and reviews are considered.

## Conclusion

By using data from the SCIE and SSCI indexes, this study conducts a comparative, dynamic, and empirical analysis focusing on the use of Web of Science and Scopus in academic papers published during 2004 and 2018. This study shows that more and more papers have used (at least mentioned) WoS/Scopus for academic research. Scopus as the new-comer, is challenging the dominating role of WoS. Although researchers from increasing number of countries/regions are involved, the main contributors are still the developed economies. China, Brazil and Iran are three developing economies who also contribute a lot to WoS- and Scopus-related research. However, their patterns in using WoS and Scopus for academic research vary significantly. The database preference may be influenced by a variety of factors including data source availability, data quality and coverage and even users' past experience. China's overrepresentation in WoS-related papers may partly due to the overemphasis of SCIE and SSCI indexed publications in China (Liu et al. 2015a, 2015b; Quan et al. 2017; Tang et al. 2015).

This study also finds a large share of WoS- and Scopus-related papers which are associated with the meta-analysis. That is to say, besides the wide use of WoS and Scopus in bibliometric related studies, both these two databases are also widely used in meta-analysis related studies, especially in China. What's more, researchers from more and more knowledge domains are using WoS and Scopus for academic research. Both the WoS and Scopus are widely used in health/medical science related domains and the traditional Information Science & Library Science field.

This short communication also has some limitations. Firstly, the search strategy used in this study is a balance between recall and precision. For example, this study only uses the topic field to identify related studies rather than searching in the full text. Many related records which may only mention the data source in the methods section will be omitted in this study. Secondly, similar to the work of Li et al. (2018), a deeper analysis focusing on the content of WoS- and Scopus-related papers is also deserved further investigation in the future. A classification of the use of these two databases into meta-analysis (as provided in Role of meta-analysis section of this paper), research evaluation and so on is also an interesting topic. The databases may be used for different purposes in different fields. Besides, this study only focuses on the use of these two databases in academic papers, the use of these two databases reflected by policy documents or evaluation practice is also deserved further investigation.

## Acknowledgements

This research is supported by the National Natural Science Foundation of China (#71801189 and #71904168) and Zhejiang Provincial Natural Science Foundation of China (#LQ18G030010 and #LQ18G010005). The authors would like to thank the referee for his/her insightful suggestions which have significantly improved the manuscript.